\documentclass[a4paper,11pt]{article}
\pdfoutput=1 

\usepackage{amsmath}
\usepackage{amssymb}
\usepackage{amsthm}
\usepackage{graphicx}
\usepackage{slashed}
\usepackage{setspace}
\usepackage{floatflt}
\usepackage{multirow}

\newcommand{\ov}{\overline}

\usepackage{color}

\usepackage{jheppub} 

\usepackage[T1]{fontenc} 

\title{String Consistency, Heavy Exotics, and the $750$ GeV \\ Diphoton Excess at the LHC: Addendum}

\author[1,2]{Mirjam Cveti{\v c},}
\author[3]{James Halverson,}
\author[4,5]{and Paul Langacker}

\affiliation[1]{Department of Physics and Astronomy,  University of Pennsylvania, \\ Philadelphia, PA 19104-6396, USA}
\affiliation[2]{Center for Applied Mathematics and Theoretical Physics,  University of Maribor, \\ Maribor, Slovenia}
\affiliation[3]{Department of Physics, Northeastern University, \\ Boston, MA 02115-5000, USA}
\affiliation[4]{School of Natural Science, Institute for Advanced Study, \\ Einstein Drive, Princeton, NJ 08540, USA}
\affiliation[5]{Department of Physics, Princeton University\\ Princeton, NJ 08544, USA}

\emailAdd{cvetic@hep.upenn.edu}
\emailAdd{j.halverson@neu.edu}
\emailAdd{pgl@ias.edu}

\abstract{We study models with heavy exotics that account for the LHC
  $750$ GeV diphoton excess in light of current vector-like quark
  bounds. Utilizing only exotics that may appear in three-stack and
  four-stack D-brane models, we show that a narrow width diphoton
  excess can be accounted for while evading existing bounds if
  multiple exotics are added, with vector-like leptons of mass
  $M_L\lesssim 375$ GeV and vector-like quarks with masses up to $\simeq
  3$ TeV. However, a large width $(\Gamma/M \sim 0.06)$, as suggested
  by the ATLAS data, cannot be easily accommodated in this
  framework. Renormalization group equations with GUT-scale boundary
  conditions show that these supersymmetric models are perturbative
  and stable. Type IIA compactifications on toroidal orbifolds allow
  for $O(10)$ Yukawa couplings in the ultraviolet.}

\notoc

\begin{document}

\begin{flushright}
\parbox[t]{1.73in}{\flushright 
UPR-1278-T}
\end{flushright}
\maketitle

\section{Introduction}

The 750 GeV diphoton excess observed by ATLAS and CMS
\cite{Seminar,CMS,ATLAS} in their $13$ TeV data has been studied in
many theoretical papers using variants on a few different types of
models. In one of the most studied types, the $750$ GeV particle is a
scalar or pseudoscalar $s$ that is produced and decays via loops of
vector-like exotic fermions \cite{Franceschini:2015kwy,McDermott:2015sck,Ellis:2015oso,Gupta:2015zzs,Martinez:2015kmn,Fichet:2015vvy,Bian:2015kjt,Falkowski:2015swt,Bai:2015nbs,Dhuria:2015ufo,Chakraborty:2015jvs,Wang:2015kuj,Murphy:2015kag,Hernandez:2015ywg,Huang:2015rkj,Badziak:2015zez,Cvetic:2015vit,Cheung:2015cug,Zhang:2015uuo,Hall:2015xds,Wang:2015omi,Salvio:2015jgu,Son:2015vfl,Cai:2015hzc,Bizot:2015qqo,Hamada:2015skp,Kang:2015roj,Jiang:2015oms,Jung:2015etr,Gu:2015lxj,Goertz:2015nkp,Ko:2016lai,Palti:2016kew,Karozas:2016hcp,Bhattacharya:2016lyg,Cao:2016udb,Faraggi:2016xnm,Han:2016bvl,Kawamura:2016idj,King:2016wep,Nomura:2016rjf,Harigaya:2016pnu,Han:2016fli,Hamada:2016vwk}. Such models have utilized a variety
of exotics, masses, and low-scale Yukawa couplings, and some have also
addressed issues of ultraviolet perturbativity and stability
\cite{Dhuria:2015ufo,Gu:2015lxj,Zhang:2015uuo,Salvio:2015jgu,Goertz:2015nkp,Hamada:2015skp,Bae:2016xni,Hamada:2016vwk}.

In our previous work \cite{Cvetic:2015vit}, we studied models of the
diphoton excess in which the addition of heavy exotics is motivated by
string consistency conditions. In type II string models these
conditions are Gauss laws for D-brane charges in the compact extra
dimensions, the D-brane tadpole cancellation conditions. They
place constraints on the chiral spectrum that go beyond
standard gauge anomaly cancellation in quantum field theory. These
additional constraints are necessary and sufficient for the
cancellation of non-abelian anomalies after D-brane pair production
\cite{Halverson:2013ska}. Most MSSM realizations in this type II
context do not satisfy the additional conditions and exotics must be
added to the theory, as studied in \cite{Cvetic:2011iq} and subsequent
works \cite{Cvetic:2012kj,Halverson:2014nwa}.  Vector-like exotics 
are also well-motivated in local F-theory models \cite{Palti:2016kew}.

In this addendum we extend our analysis, simultaneously taking into account
current bounds on vector-like quarks, perturbativity and stability of the models up
to a high scale, and the decay rates necessary to account for the diphoton excess.

\section{Renormalization Group Equations and Infrared Fixed Points}

We consider models where the $750$ GeV
particle is a scalar degree of freedom in a chiral supermultiplet $S$
that couples to $N_i$ exotic vector-like chiral multiplets $X_i,\ov X_i$ in the superpotential
via a Yukawa coupling $\gamma_i\, S X_i \ov X_i$. The exotics $X_i$ and $\ov X_i$ transform
as $(n_3^i, n_2^i)_{y_i}$ and  $(n^{i\ast}_3, n^{i\ast}_2)_{-y_i}$, respectively, with $q_i=t_{3,i}+y_i$.

In supersymmetric models the gauge couplings are governed by the
renormalization group equations
\begin{align}
16 \pi^2 \beta_{g_3} & = \left[ -3 + 2 \sum_i N_i T^3(i) n^i_2 \right] g_3^3 \nonumber \\
16 \pi^2 \beta_{g_2} & = \left[ +1 +2 \sum_i N_i T^2(i) n^i_3 \right] g_2^3 \nonumber \\
16 \pi^2 \beta_{g_1} & = \left[ +\frac{33}{5} + 2 \sum_i N_i T^1(i) n^i_2 n^i_3 \right] g_1^3 .
\end{align}
where $\beta_{g_i} \equiv dg_i/dt$ with $t=\ln(\mu/\mu_0)$. $T^a(i)$ is
the Dynkin index for representation $i$ in group $a$. The Dynkin
indices of low-dimensional representations are $T^3(i) = (3,
1/2, 0)$ for $n^i_3=(8, 3, 1)$, $T^2(i) = (2, 1/2, 0)$ for $n^i_2=(3,
2, 1)$, and $T^1(i) \equiv \frac{3}{5} y_i^2$.   We have
used the GUT-normalized gauge coupling $g_1$ for $U(1)_Y$, which is
related to th ordinary $g'$ by $g_1=\sqrt{5/3} g'$. The beta functions for
the Yukawa coupling $\gamma_i$ are [see, e.g., \cite{Martin:1997ns}]
\begin{equation}
16 \pi^2 \beta_{\gamma_i} = 2 \gamma_i |\gamma_i|^2 +  \gamma_i \left( \sum_j N_j n^j_2 n^j_3 |\gamma_j|^2\right) 
 -4 \gamma_i \sum_{a=1}^3 C_2^a(i) g_a^2.
\end{equation}

The specific set of models we will study have the MSSM spectrum with three right-handed neutrinos
augmented by $N_Q$ $(3,2)_{1/6}+(\ov 3,2)_{-1/6}$ pairs, $N_U$ $(3,1)_{2/3}+(\ov 3,1)_{-2/3}$ pairs,
$N_D$ $(3,1)_{-1/3}+(\ov 3, 1)_{1/3}$ pairs, $N_L$ $(1,2)_{-1/2}+(1,2)_{1/2}$ pairs, and $N_E$ $(1,1)_1+(1,1)_{-1}$
pairs, all of which occur frequently in the type IIA compactifications \cite{Cvetic:2015vit,Cvetic:2011iq}. The subscripts denote that one of the chiral multiplets in the pair has the same MSSM quantum numbers as
the associated MSSM superfield, e.g., $Q, U, D, L, E$. The Yukawa couplings in this model
will be labeled similarly, i.e., $\gamma_Q, \gamma_U, \gamma_D, \gamma_L, \gamma_E$. The beta functions
for the gauge couplings are
\begin{align}
16 \pi^2 \beta_{g_3} &= g_3^3(-3+2N_Q+N_U+N_D) \nonumber \\
16 \pi^2 \beta_{g_2} &= g_2^3(1+3N_Q+N_L) \nonumber \\
16 \pi^2 \beta_{g_1} &= g_1^3\left(\frac{33}{5} + \frac65 \left(\frac{N_Q}{6}+\frac{4N_U}{3}+\frac{N_D}{3}+\frac{N_L}{2}+N_E\right) \right). 
\end{align}
Those for the Yukawa couplings are
\begin{align}
16 \pi^2 \beta_{\gamma_Q} &= \gamma_Q \left[2|\gamma_Q|^2 + \alpha -4\left(\frac43 g_3^2+\frac34 g_2^2+\frac35\left(\frac16\right)^2g_1^2\right) \right] \nonumber \\
16 \pi^2 \beta_{\gamma_U} &= \gamma_U \left[2|\gamma_U|^2 + \alpha -4\left(\frac43 g_3^2+\frac35 \left(\frac23\right)^2g_1^2\right) \right] \nonumber \\
16 \pi^2 \beta_{\gamma_D} &= \gamma_D \left[2|\gamma_D|^2 + \alpha -4\left(\frac43 g_3^2+\frac35 \left(\frac13\right)^2g_1^2\right) \right] \nonumber \\
16 \pi^2 \beta_{\gamma_L} &= \gamma_L \left[2|\gamma_L|^2 + \alpha -4\left(\frac34 g_2^2+\frac35 \left(\frac12\right)^2g_1^2\right) \right] \nonumber \\
16 \pi^2 \beta_{\gamma_E} &= \gamma_E \left[2|\gamma_E|^2 + \alpha -4\left(\frac35 g_1^2\right) \right], \label{betas}
\end{align}
where 
\begin{equation}
\alpha = 6N_Q|\gamma_Q|^2+3N_U|\gamma_U|^2+3N_D|\gamma_D|^2+2N_L|\gamma_L|^2+N_E|\gamma_E|^2.
\end{equation}
We will study models with specific values for the tuple $(N_Q,N_U,N_D,N_L,N_E)$.

\vspace{1cm}

The perturbative nature of specific models is ensured in part by the
existence of infrared fixed points. For reasonable
ultraviolet boundary conditions for the Yukawa couplings in the range
$[0.1,10]$, Yukawa couplings of vector-like quarks
often approach their fixed points. 

Let us first justify this range of 
UV Yukawa couplings in type IIA compactifications with intersecting D6-branes, which are one of the contexts
for our previous work \cite{Cvetic:2015vit}. We shall focus on the allowed
magnitudes of the Yukawa couplings at the string scale.  These were
calculated exactly at the string (world-sheet) tree level for toroidal
compactifications in \cite{Cvetic:2003ch}.  The full
expression (both classical and quantum part of the string tree level
amplitude) for branes wrapping factorizable three-cycles of $T^6=T^2\times
T^2\times T^2$ is written as (see, eq. (3) of \cite{Cvetic:2003ch}):
\begin{equation}
\gamma=\sqrt{2}g_s 2\pi\prod_{j=1}^3\left[\frac{16\pi^2 B(\nu_j,1-\nu_j)}{
    B(\nu_j,\lambda_j)B(\nu_j,1-\nu_j-\lambda_j)}\right]^\frac14\sum_m\exp\left(-\frac{A_j(m)}{2\pi\alpha'}\,\right)  .
    \end{equation}
Here the chiral superfields are localized at the intersections of pairs of D6-branes, which 
intersect at respective angles $\pi\nu_j$, $\pi \lambda_j$,  and $\pi- \pi\nu_j-\pi\lambda_j$ on the   $j$th two-torus.
$A_j(m)$ is the area of the triangle formed by the three
    intersecting D6-branes on the j-th two-torus and  $g_s=e^{\Phi/2}$, with $\Phi$ corresponding to the Type IIA dilaton.  The beta function $B(p,q)$ is defined in terms of $\Gamma(p)$ functions as $B(p,q)\equiv \frac{\Gamma(p)\, \Gamma(q)}{\Gamma(p+q)}$.
The coupling is between two fermion  fields and a scalar field, i.e., the
massless states appearing at the respective intersections,  whose
kinetic energies are taken to be canonically normalized.

\begin{figure}[tb]
\begin{center}
\includegraphics[scale=.6]{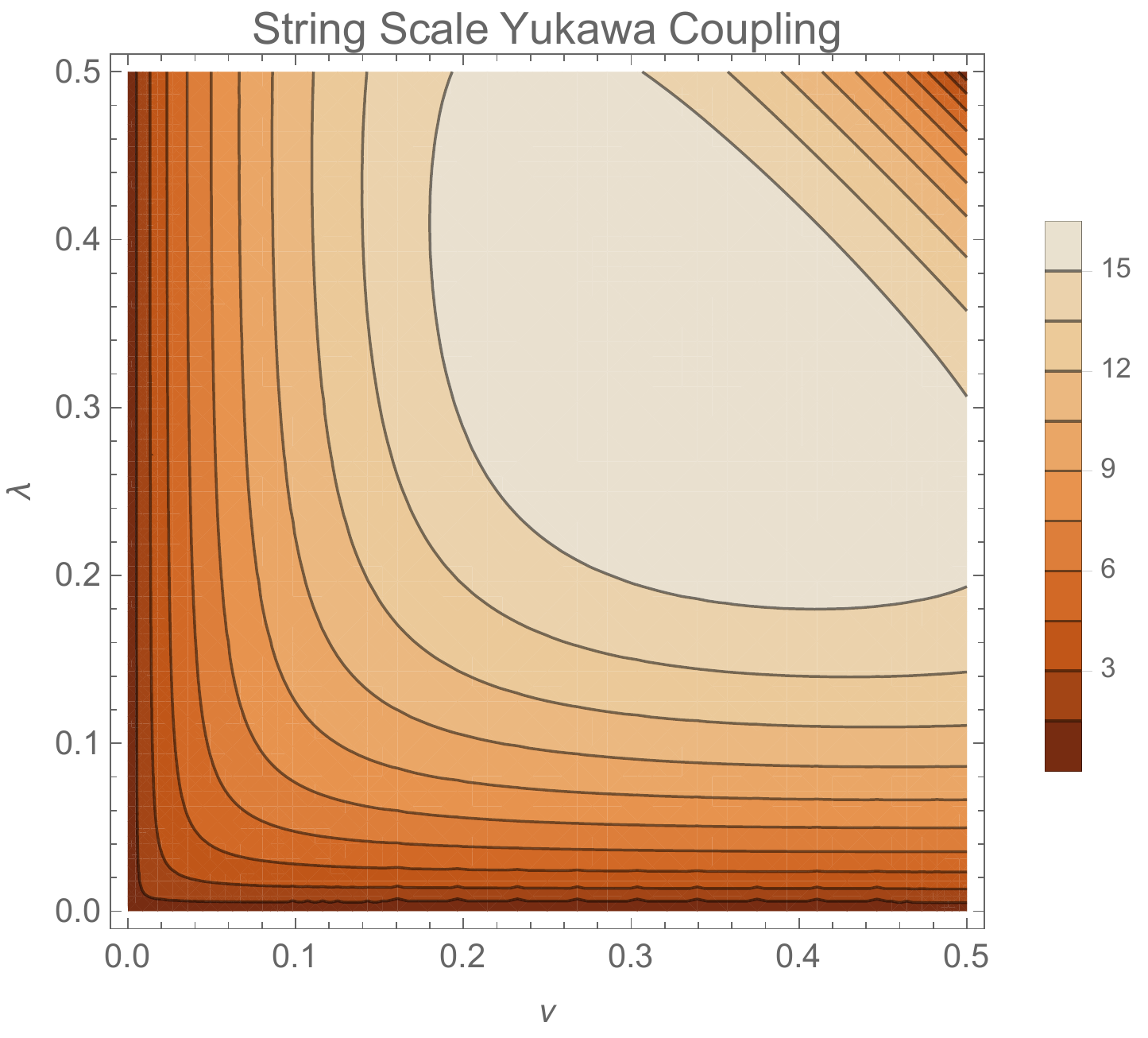}
\end{center}
\caption{String scale Yukawa coupling for $g_s=0.2$ and brane angles $\lambda, \nu$.}
\label{fig:UVYukawa}
\end{figure}

The magnitude of Yukawa couplings can be surprisingly
sizable, reaching ${\cal O}(10)$, while having the string coupling
still perturbative. The choice of brane angles and disc instanton
areas that maximize the Yukawa coupling are
$\nu_j=\lambda_j=1-\nu_j-\lambda_j=\frac{1}{3}$ ($j=1,2,3$) and
$A_{j}(m=1) =0$, respectively. Taking these values and a perturbative
string coupling $g_s=0.2$, the Yukawa coupling is $\gamma=17$. Since
$\gamma\propto g_s$ the Yukawa coupling can be even higher while
remaining within a perturbative string framework. Taking universal
angles for all tori $\nu_j=:\nu$ and $\lambda_j=:\lambda$, the
dependence of $\gamma$ on these angles is presented in Figure
\ref{fig:UVYukawa}, which demonstrates that for zero disc instanton
area the Yukawa couplings $\gamma$ are $>1$ for a wide variety of angles.
In summary, perturbative Type IIA string theory therefore allows for a
range of $\gamma$ including large values in the interval $\gamma\in [{\cal O}(1),\ {\cal
  O}(50)]$.

We now turn to an approximate analytic analysis of the range of Yukawa
couplings in the IR regime, in particular for $\gamma_Q$.  We will
demonstrate that for a broad range of UV boundary conditions
$\gamma_Q$ robustly approaches an approximate IR fixed point, governed
by the IR value of $g_3$, the largest gauge coupling.  (Note, for
example, an early analysis of such an IR behavior for the fourth
family Yukawa couplings within the MSSM \cite{Cvetic:1985fp}.)

First one observes from (\ref{betas})  that $\gamma_L$  tends to decrease in the IR regime due to a positive, dominant contribution from $\gamma_Q$ and a smaller, negative contributions from $g_2$. We shall reconfirm post-factum that  for $\gamma_L ={\cal O}(1)$ in the UV,  $\gamma_L < \gamma_Q$ in the IR.  

To illustrate the IR fixed point behavior, let us study the beta
function for $\gamma_Q$ in (\ref{betas}) in the case of only $Q$
exotics, i.e., $N_Q\ne 0$ and $N_{U,D,L,E}=0$. For simplicity we
neglect $g_{1,2}$ relative to $g_3$ and replace the running $g_{3}$
with its (approximately ``constant'') IR value at $\Lambda_{IR}\sim 1$
TeV. This approximation is justified since the gauge couplings run
logarithmically with the scale $\Lambda$, while the IR fixed point for
Yukawa couplings is approached with a power-law for $\Lambda$.  With
these approximations we obtain:
\begin{equation}
16 \pi^2 \beta_{\gamma_Q} = \gamma_Q \left[2(1+3N_Q)\gamma_Q^2 - \frac{16}{3} g_{3\, IR}^2 \right] \, ,
\end{equation} 
which is easily solved to yield
\begin{equation}
\gamma_{Q\, IR}^2= \frac{a}
{\left[1-\left(1-\frac{a}{\gamma_{Q\, UV}^2}\right) \left(\frac{\Lambda_{IR}}{\Lambda_{UV}}\right)^{\frac{2a}{b}}\right]},
\label{eqn:solvedyukawa}
\end{equation}
where $a= \frac{16}{3} \frac{g_{3\, IR}^2}{2(1+3N_Q)}$ and $b=\frac{16\pi^2}{2(1+3N_Q)}$.
We thus observe an IR  robust fixed point governed  by $a$. Although \eqref{eqn:solvedyukawa} gives a reasonable
approximation to $\gamma_Q$, we will use the exact solutions to \eqref{betas} in our subsequent
analysis.

\section{Perturbative and Stable Models of the Diphoton Excess}

\begin{figure}[tb]
\begin{center}
\includegraphics[scale=.3]{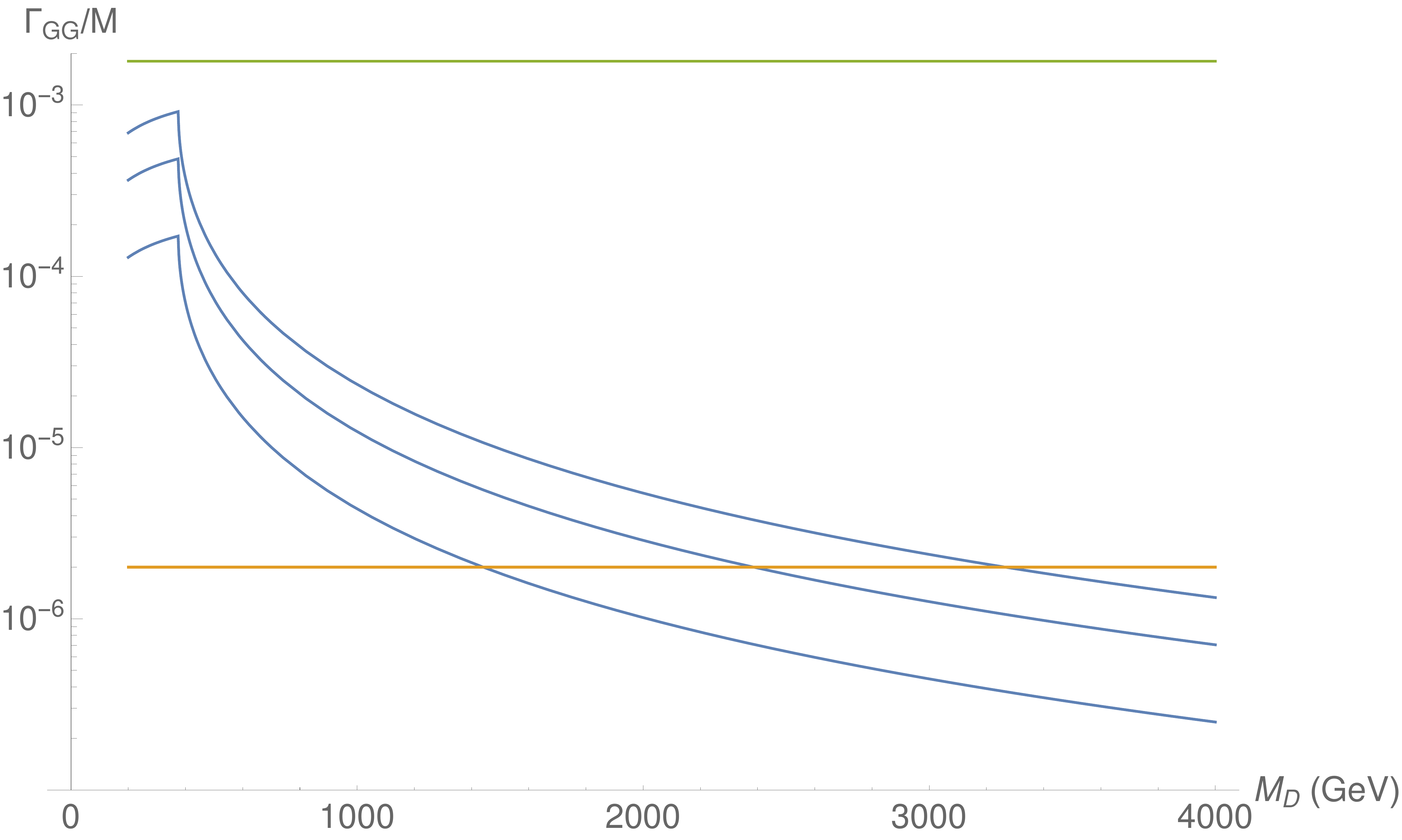}
\end{center}
\caption{$\Gamma_{GG}/M$ as a function of $M_D$ for $N_D=N_L=3,2,1$ are the top, middle, and bottom blue lines, respectively. Upper
and lower bounds on the rate from \cite{Franceschini:2015kwy} are also given.}
\label{fig:GamGG}
\end{figure}

The existing experimental bounds on new vector-like fermions are very
model dependent. Assuming decays into standard model particles such as
$D\rightarrow W t, Z b,$ or $ Hb$ the current 95\% C.L. lower limits
are in the range 740-900 GeV~\cite{Khachatryan:2015gza} or 575-813
GeV~\cite{Aad:2015kqa} for CMS and ATLAS, respectively.  The
corresponding limits for a heavy charge-2/3 quark are 720-920
GeV~\cite{Khachatryan:2015oba} and 715-950 GeV~\cite{Aad:2015kqa}.
Those for charged and neutral leptons are much weaker, typically
around 100 GeV~\cite{Agashe:2014kda}, although some mass ranges up to
$\sim$180 GeV are excluded \cite{Aad:2015dha}. We will simply assume
$M_f \gtrsim 750$ GeV (quarks) and $\gtrsim 200$ GeV (leptons).

\begin{figure}[htb]
\begin{center}
\makebox[\textwidth][c]{
\includegraphics[scale=.4]{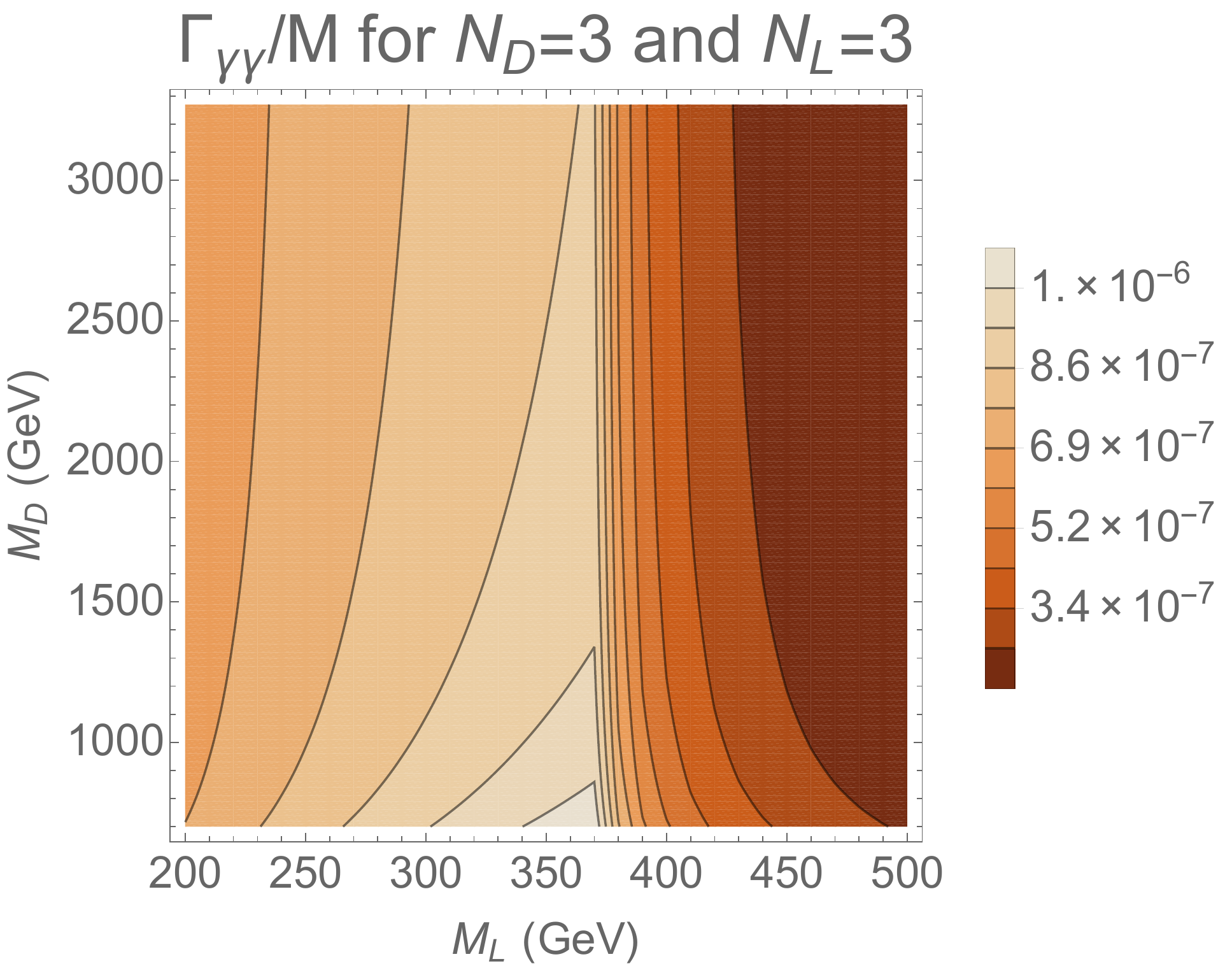}
\includegraphics[scale=.4]{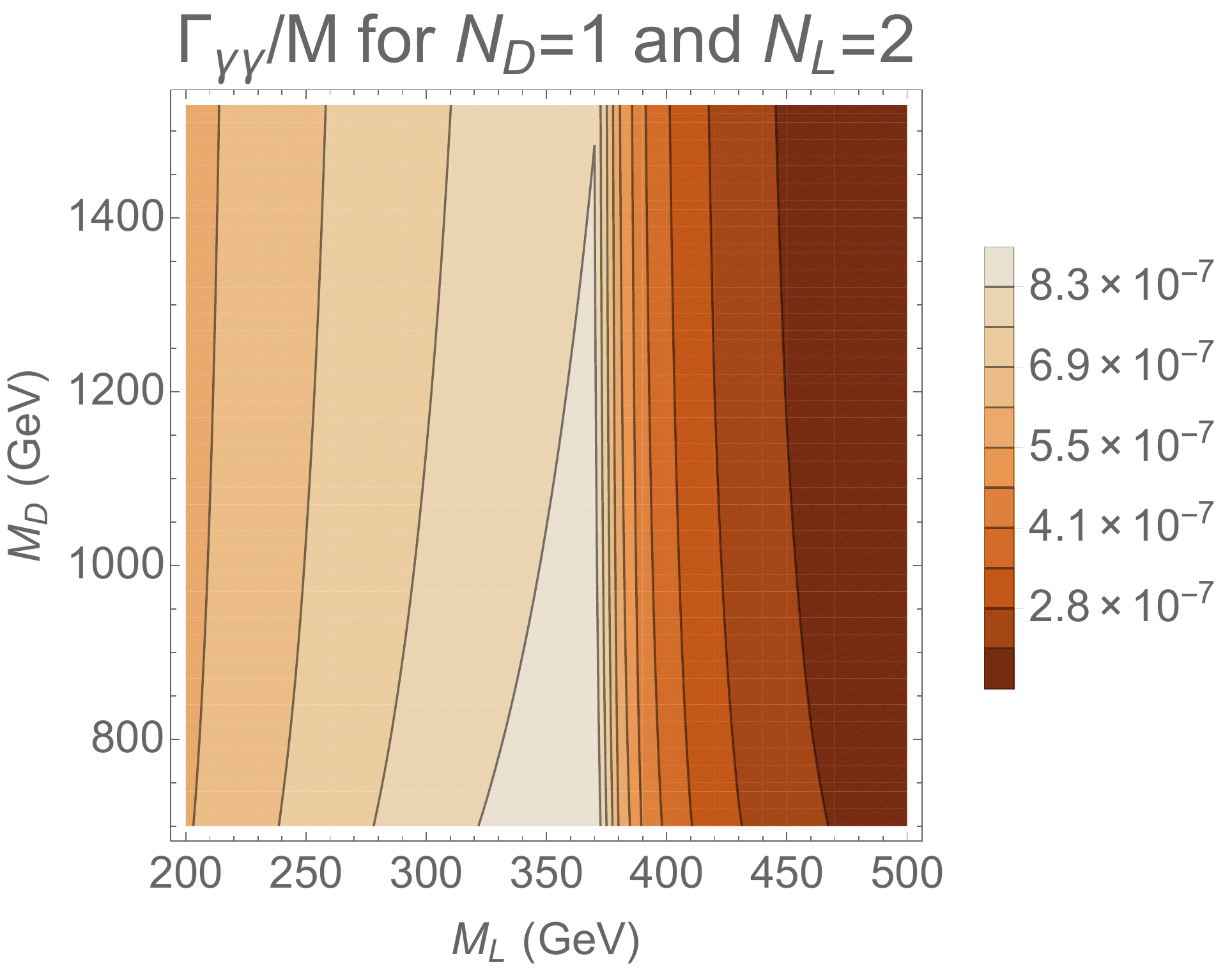}
}
\end{center}
\caption{The partial width into photons for models with $(N_D,N_L)=(3,3)$ and $(1,2)$.}
\label{fig:33and12}
\end{figure}

We consider models with $N_L$ vector-like lepton doublets and $N_D$ down-type
quark chiral supermultiplets that couple in the superpotential to an
MSSM singlet $S$ that contains a pseudoscalar $s$ of mass $M_s=750$
GeV. We choose $t_{UV}=30$ $(t_{IR}=0)$ relative to a reference scale
$\mu_0=750$ GeV, corresponding to $\mu_{UV}=8.0\times 10^{15}$ GeV
($\mu_{IR}=750$ GeV).  We assume universal ultraviolet Yukawa
couplings with $\gamma_{UV}=1$ and universal masses $M_L$ and $M_D$
for the vector-like leptons and quarks. The total decay width of $s$ is 
$\Gamma_{tot}=\Gamma_{GG}+\Gamma_{\gamma\gamma}$,
where the expressions for partial widths are given in \cite{Franceschini:2015kwy}.

The partial width $\Gamma(s\to GG)$ must satisfy $\Gamma_{GG}/M\geq
8\times 10^{-7}$ to be in the preferred blue band on the left side of
Figure $1$ in \cite{Franceschini:2015kwy}.  We study
$\Gamma_{GG}/M\geq 2\times 10^{-6}$, for which the blue band flattens
out and the analysis of the rate into two photons is simplified. For
$N_D=N_L=3,2,1$ the rate $\Gamma_{GG}$ is computed in Figure
\ref{fig:GamGG} as a function of the exotic quark mass
$M_D$. $\Gamma_{GG}/M\geq 2\times 10^{-6}$ for $M_5\lesssim 3270,
2380, 1430$ GeV, respectively, well within vector-like quark
bounds. These exotic representations embed into $5 + 5^\ast$ pairs,
which maintain MSSM-like gauge unification to lowest order, and
$N_D=N_L=3$ is motivated by $E_6$ models.

For $\Gamma_{GG}/M \geq 2\times 10^{-6}$, the partial width into
photons must satisfy \cite{Franceschini:2015kwy} $6\times
10^{-7}\lesssim \Gamma_{\gamma\gamma}/M \lesssim 2\times10^{-6}$ at
2$\sigma$, assuming no other contributions to the width. The $E_6$
motivated model $(N_D,N_L)=(3,3)$ and the minimal model that can
account for the data $(N_D,N_L)=(1,2)$ are presented in Figure
\ref{fig:33and12}.  In each plot $M_D$ goes up to the maximal value
that allows for $\Gamma_{GG}/M\geq 2\times10^{-6}$. In both cases
obtaining a large enough $\Gamma_{\gamma\gamma}/M$ requires
$M_L\lesssim 375$ GeV, while obtaining a large enough $\Gamma_{GG}/M$
requires $M_D\lesssim 3.3, 1.6$ TeV, respectively.

Similar $\Gamma_{\gamma\gamma}/M$ plots for all $N_D,N_L\in \{1,2,3\}$ are presented in Figures \ref{fig:nd3}, \ref{fig:nd2}, and \ref{fig:nd1}.
Interestingly, $\Gamma_{\gamma\gamma}/M$ tends to increase with decreasing $N_D$ for fixed $N_L$.

\acknowledgments 
This research is supported in part by the DOE Grant
Award DE-SC0013528, (M.C.), the Fay R. and Eugene L. Langberg Endowed
Chair (M.C.) and the Slovenian Research Agency (ARRS) (M.C.). 
J.H. is supported by startup funding from Northeastern University.

\appendix

\begin{figure}[htb]
\begin{center}
\makebox[\textwidth][c]{
\includegraphics[scale=.4]{dl33.pdf}
\includegraphics[scale=.4]{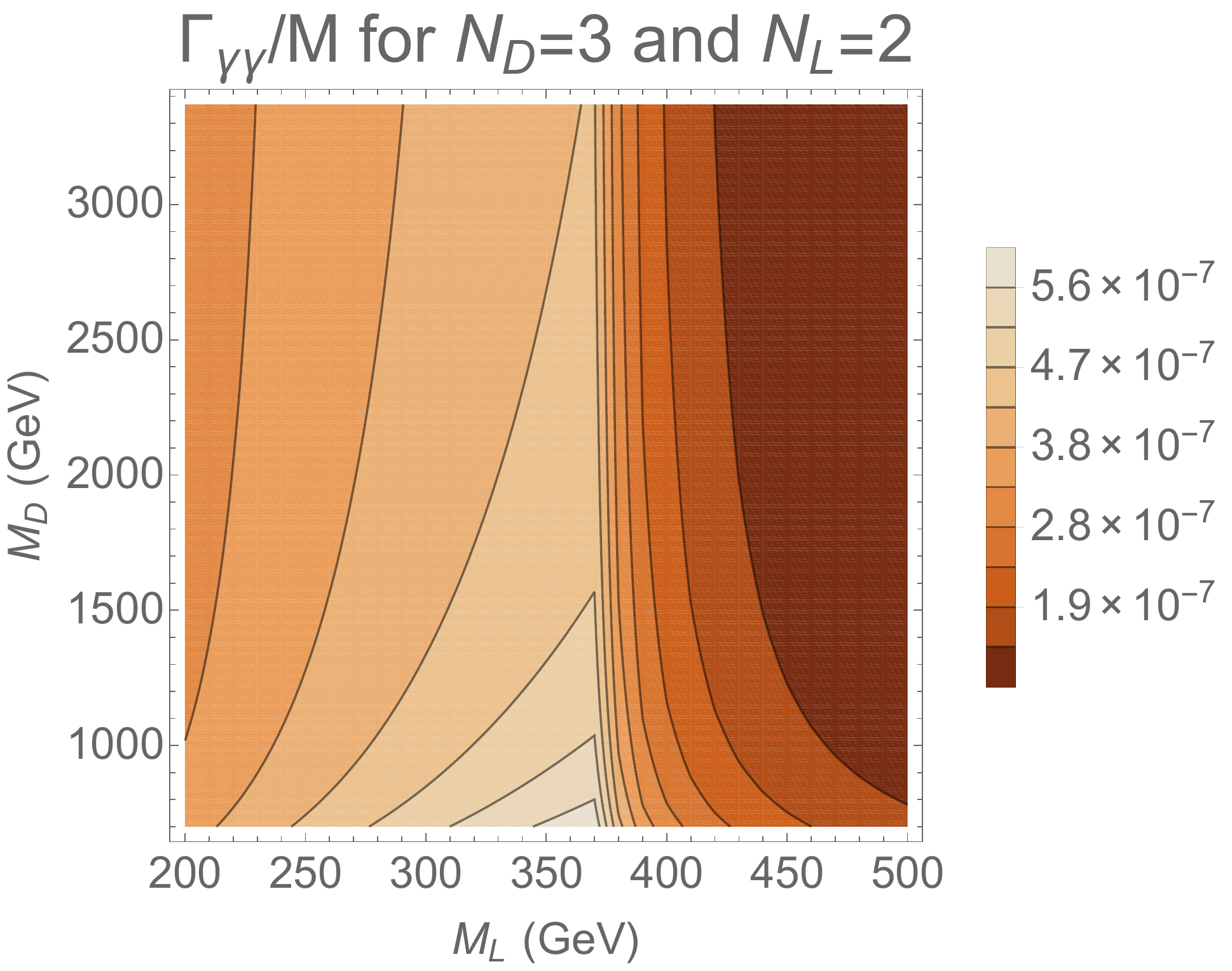}
}
\makebox[\textwidth][c]{
\includegraphics[scale=.4]{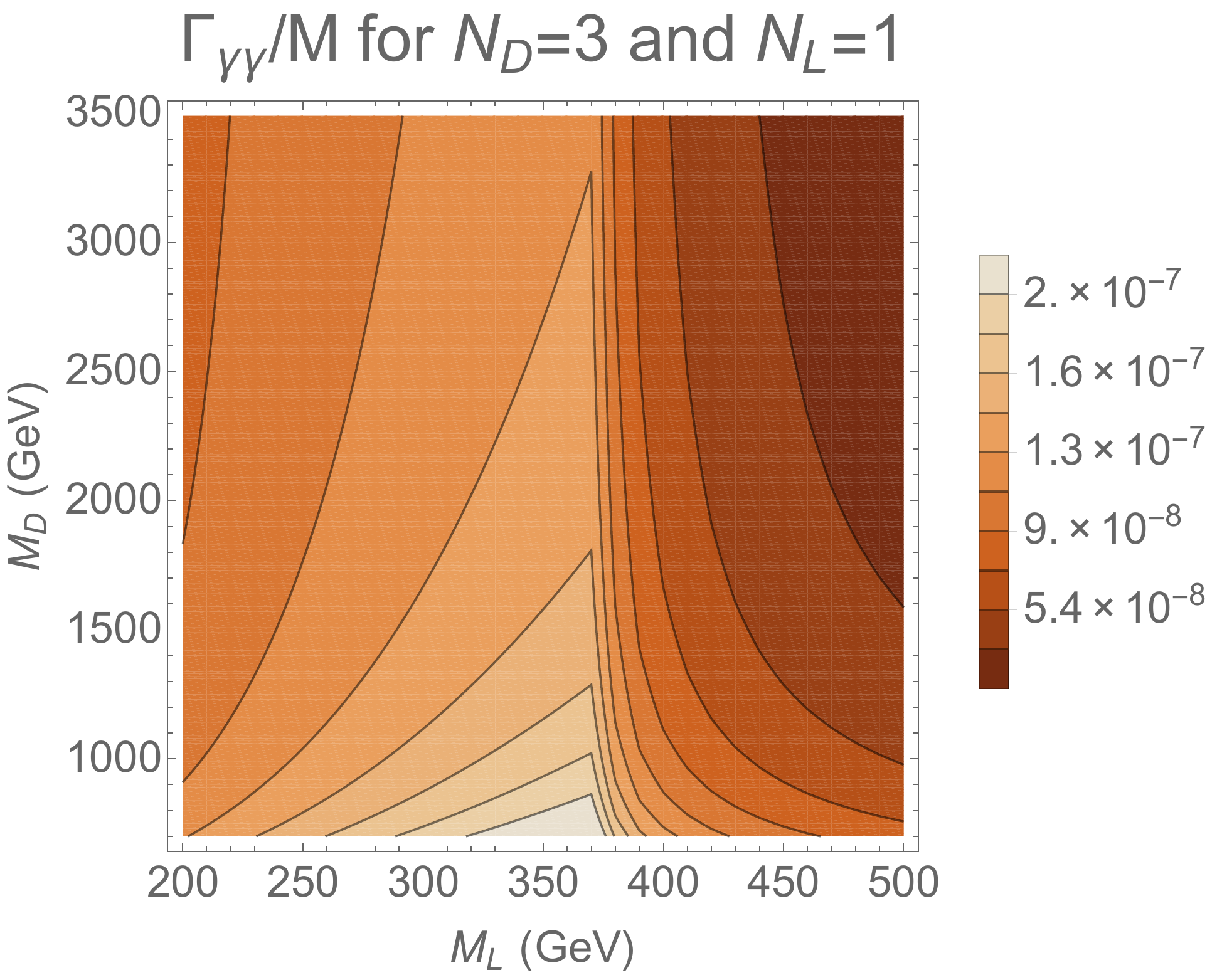}
}
\end{center}
\caption{The partial width into photons for models with $N_D=3$ and $N_L=3,2,1$.}
\label{fig:nd3}
\end{figure}

\begin{figure}[htb]
\begin{center}
\makebox[\textwidth][c]{
\includegraphics[scale=.4]{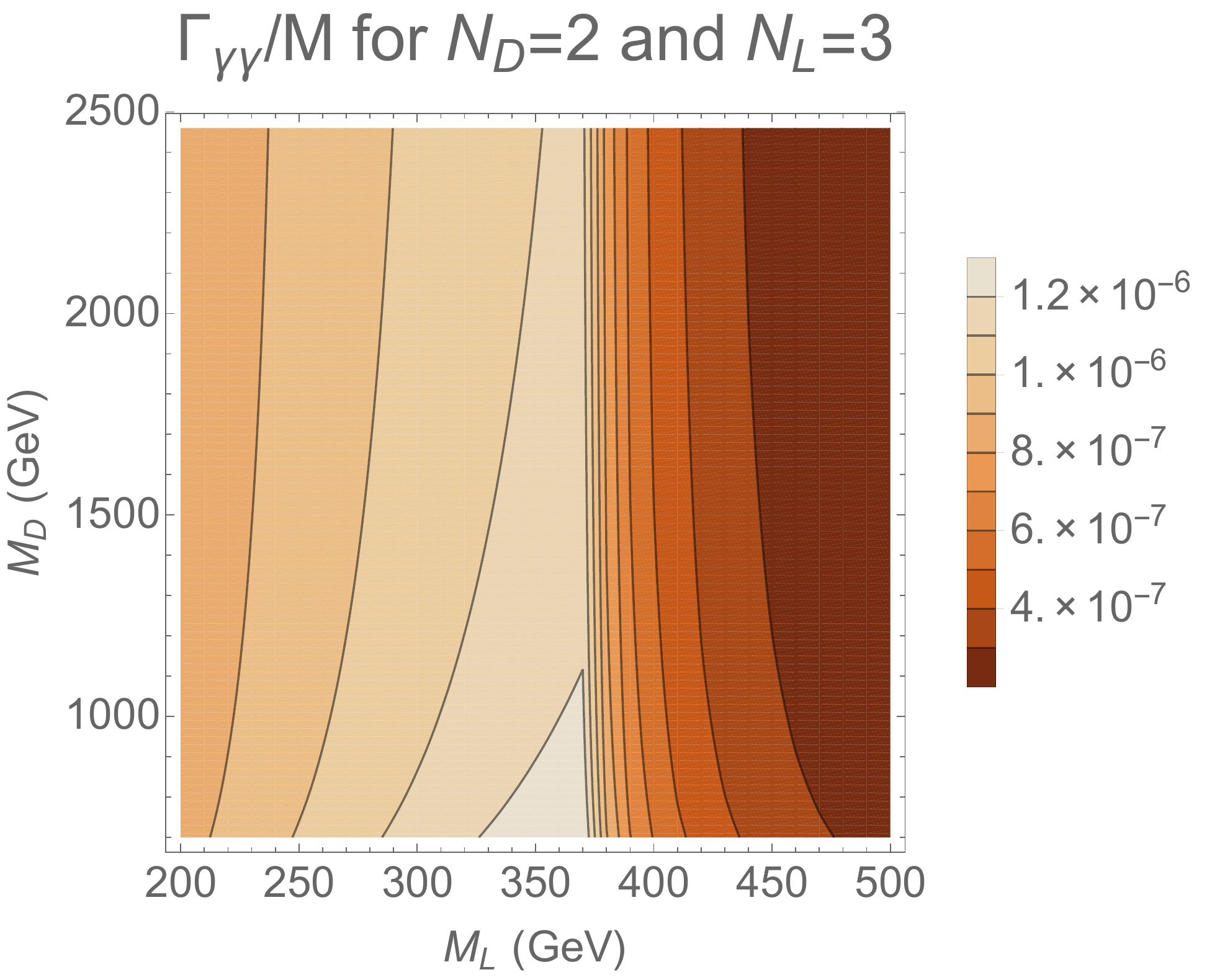}
\includegraphics[scale=.4]{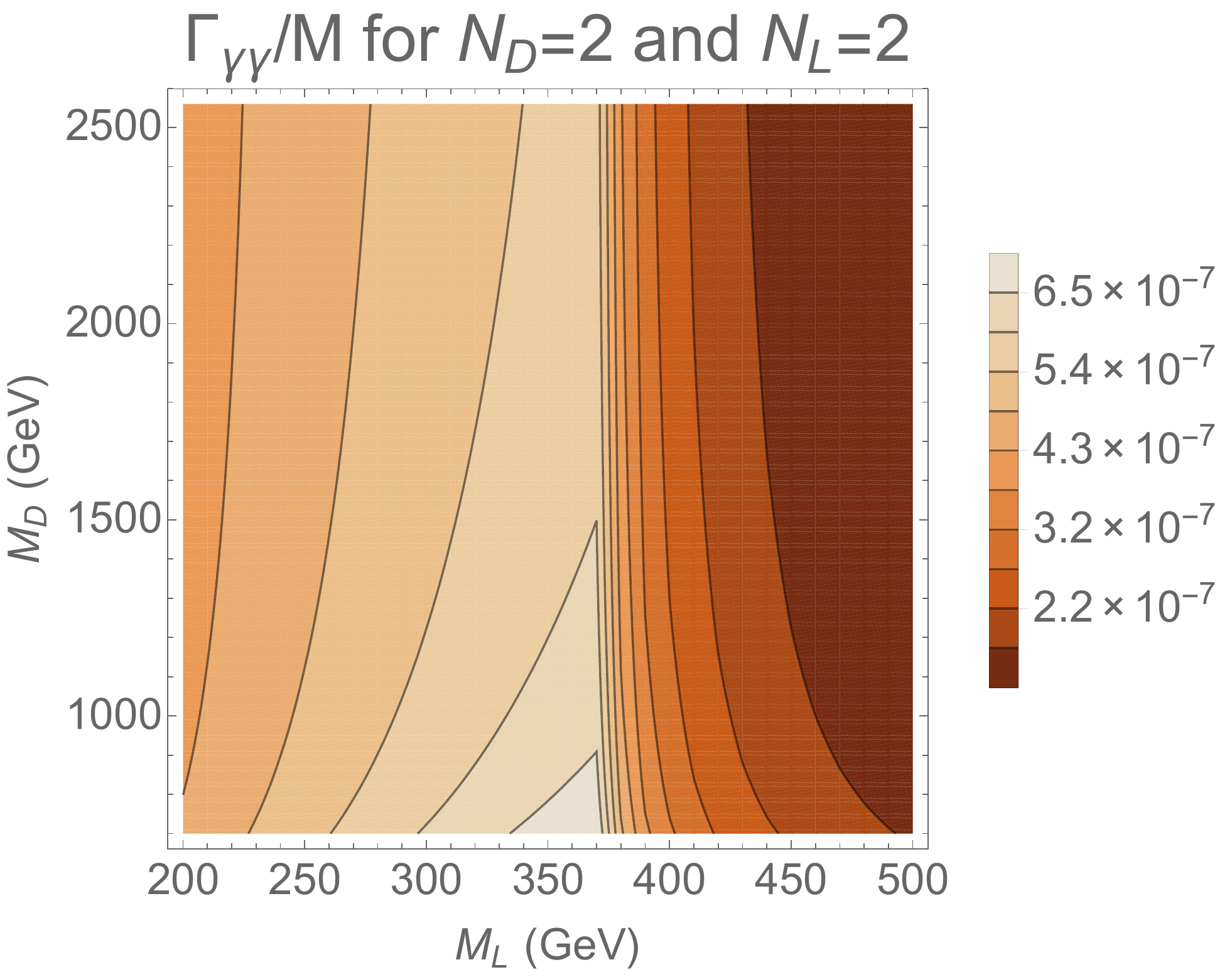}
}
\makebox[\textwidth][c]{
\includegraphics[scale=.4]{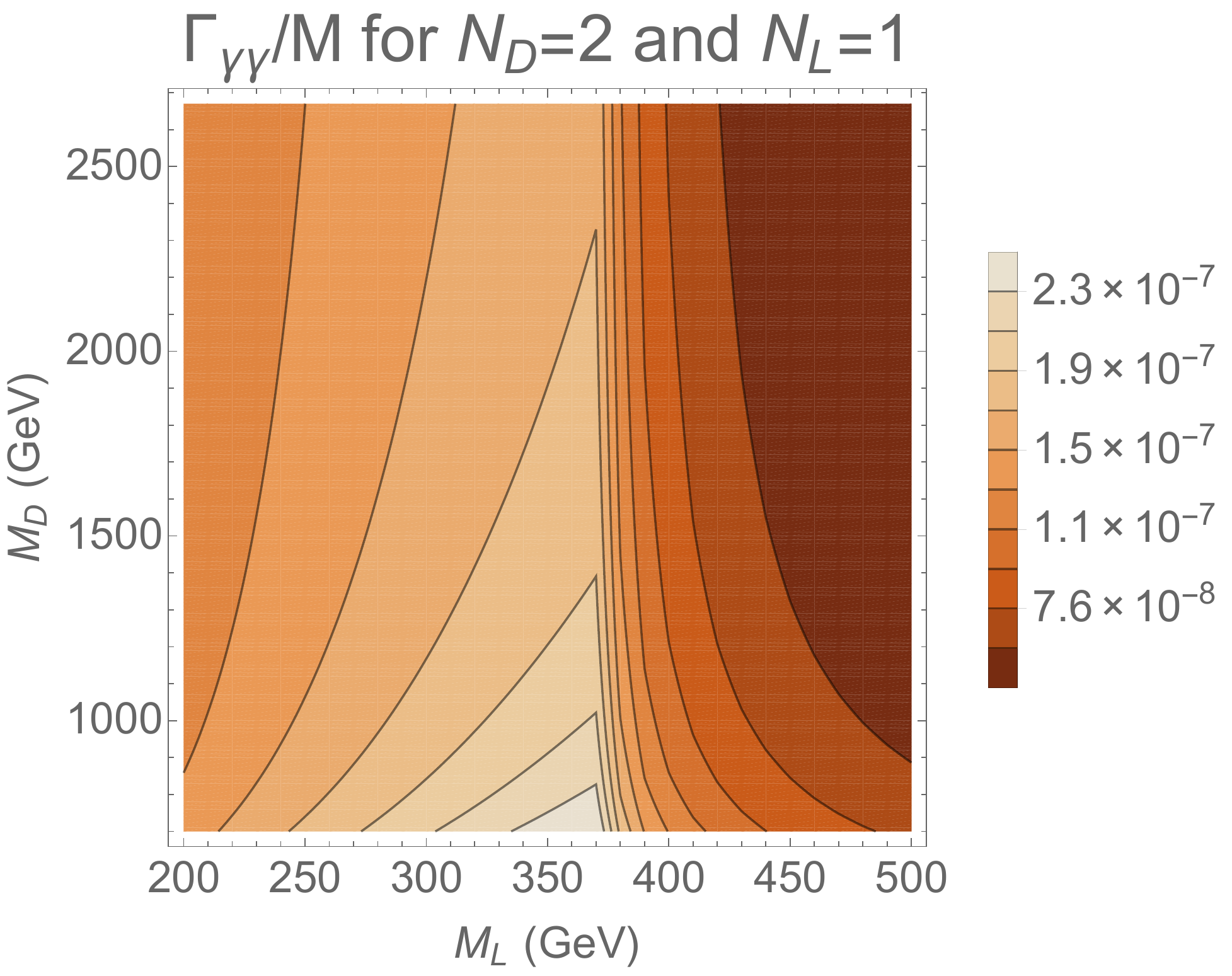}
}
\end{center}
\caption{The partial width into photons for models with $N_D=2$ and $N_L=3,2,1$.}
\label{fig:nd2}
\end{figure}

\begin{figure}[htb]
\begin{center}
\makebox[\textwidth][c]{
\includegraphics[scale=.4]{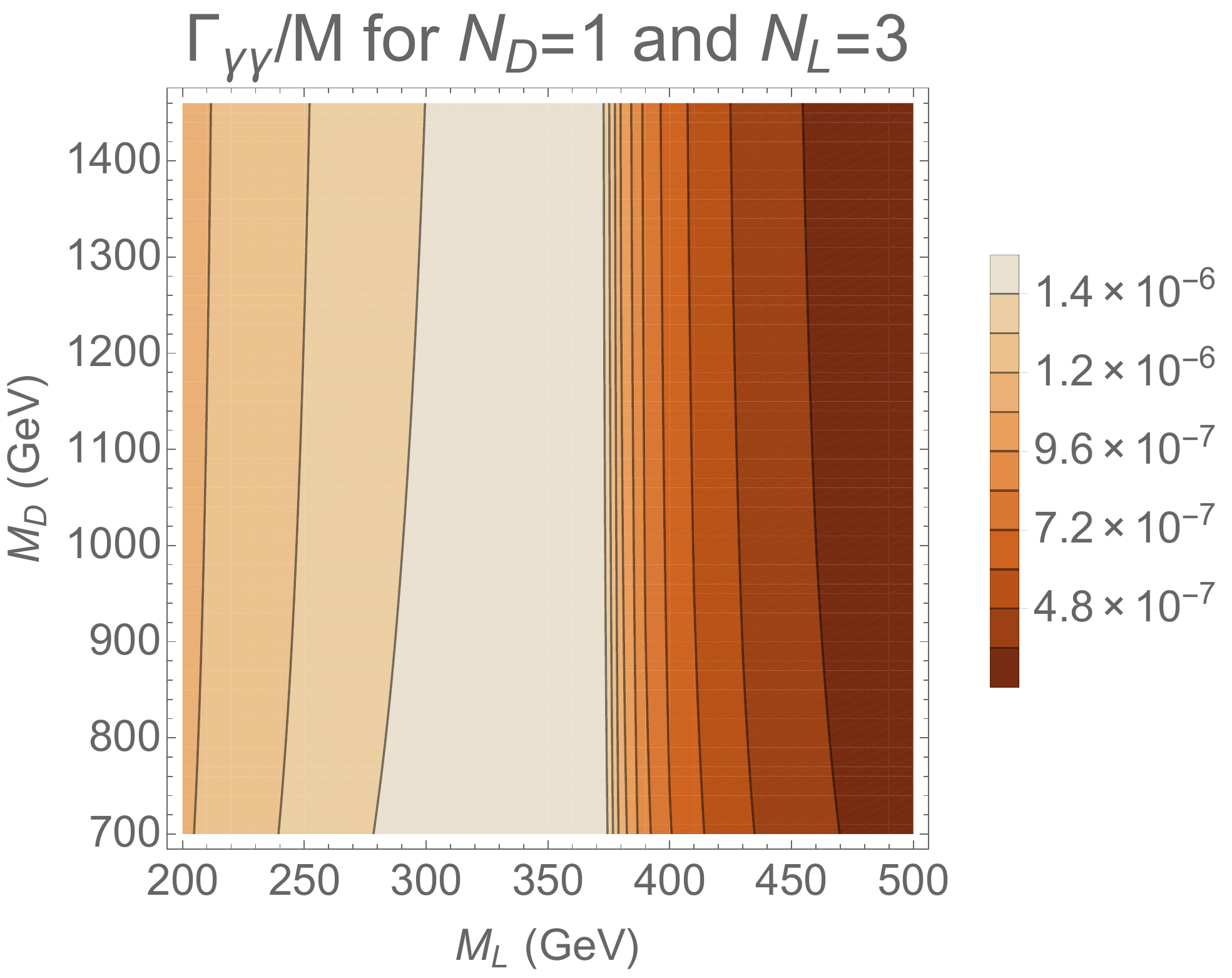}
\includegraphics[scale=.4]{dl12.pdf}
}
\makebox[\textwidth][c]{
\includegraphics[scale=.4]{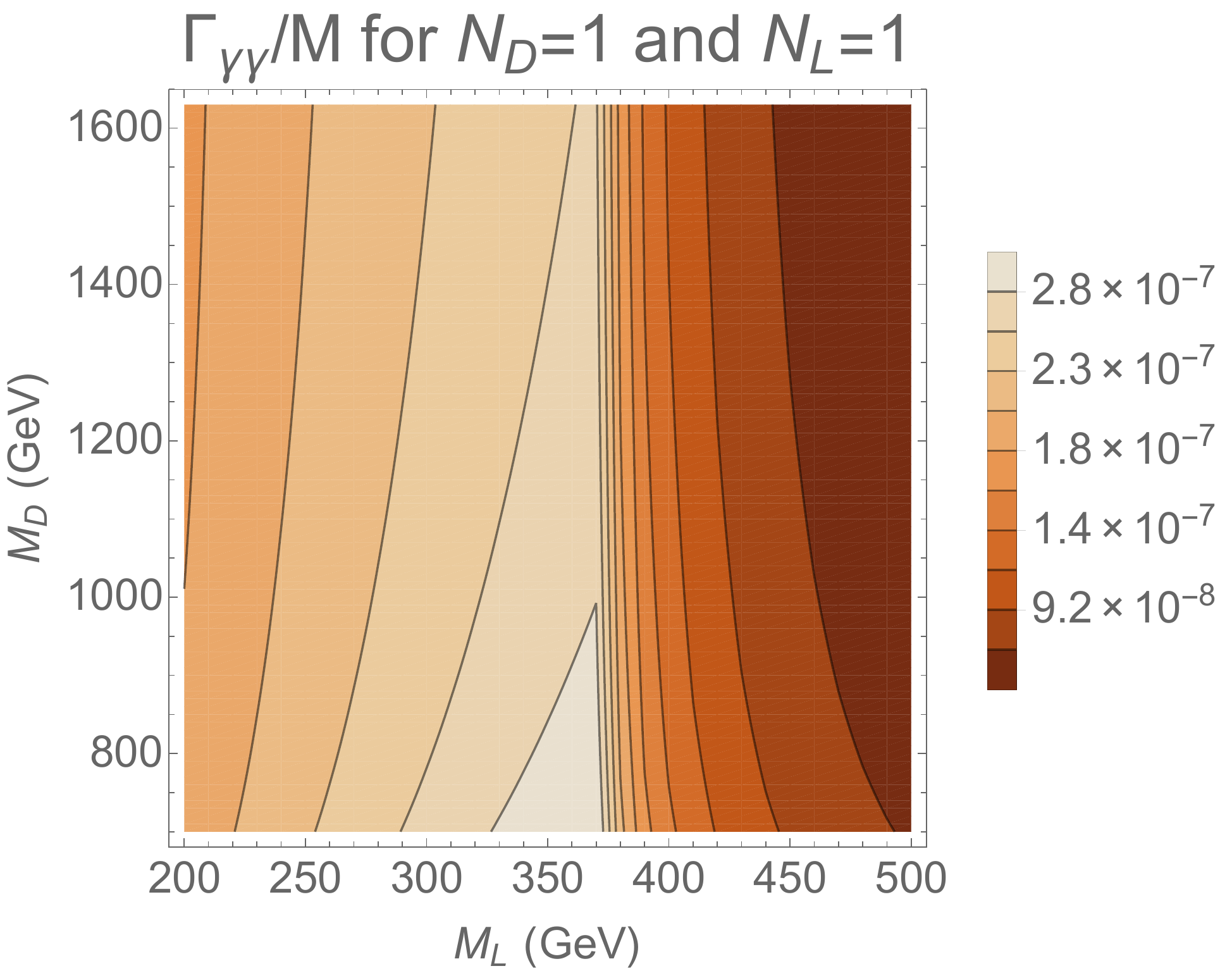}
}
\end{center}
\caption{The partial width into photons for models with $N_D=1$ and $N_L=3,2,1$.}
\label{fig:nd1}
\end{figure}

\bibliographystyle{JHEP}
\bibliography{refs}

\providecommand{\href}[2]{#2}\begingroup\raggedright\begin{thebibliography}{10}

\bibitem{Seminar}
{\it { Talks by Jim Olsen and Marumi Kado, CERN, 15 Dec. 2015.}},  {\em LHC
  seminar \href{http://indico.cern.ch/event/442432/}{{\em ATLAS and CMS physics
  results from Run 2}}}.

\bibitem{CMS}
{\it {CMS note, Search for new physics in high mass diphoton events in
  proton-proton collisions at $13$ TeV}},  {\em CMS PAS EXO-15-004}.

\bibitem{ATLAS}
{\it {ATLAS note, \em Search for resonances decaying to photon pairs in 3.2
  fb$^{-1}$ of $pp$ collisions at $\sqrt{s}=13$ TeV with the ATLAS detector}},
  {\em ATLAS-CONF-2015-081}.

\bibitem{Franceschini:2015kwy}
R.~Franceschini, G.~F. Giudice, J.~F. Kamenik, M.~McCullough, A.~Pomarol,
  R.~Rattazzi, M.~Redi, F.~Riva, A.~Strumia, and R.~Torre, {\it {What is the
  gamma gamma resonance at 750 GeV?}},
  \href{http://arxiv.org/abs/1512.04933}{{\tt arXiv:1512.04933}}.

\bibitem{McDermott:2015sck}
S.~D. McDermott, P.~Meade, and H.~Ramani, {\it {Singlet Scalar Resonances and
  the Diphoton Excess}},  \href{http://arxiv.org/abs/1512.05326}{{\tt
  arXiv:1512.05326}}.

\bibitem{Ellis:2015oso}
J.~Ellis, S.~A.~R. Ellis, J.~Quevillon, V.~Sanz, and T.~You, {\it {On the
  Interpretation of a Possible $\sim 750$ GeV Particle Decaying into $\gamma
  \gamma$}},  \href{http://arxiv.org/abs/1512.05327}{{\tt arXiv:1512.05327}}.

\bibitem{Gupta:2015zzs}
R.~S. Gupta, S.~J{\" a}ger, Y.~Kats, G.~Perez, and E.~Stamou, {\it
  {Interpreting a 750 GeV Diphoton Resonance}},
  \href{http://arxiv.org/abs/1512.05332}{{\tt arXiv:1512.05332}}.

\bibitem{Martinez:2015kmn}
R.~Martinez, F.~Ochoa, and C.~F. Sierra, {\it {Diphoton decay for a $750$ GeV
  scalar boson in an $U(1)'$ model}},
  \href{http://arxiv.org/abs/1512.05617}{{\tt arXiv:1512.05617}}.

\bibitem{Fichet:2015vvy}
S.~Fichet, G.~von Gersdorff, and C.~Royon, {\it {Scattering Light by Light at
  750 GeV at the LHC}},  \href{http://arxiv.org/abs/1512.05751}{{\tt
  arXiv:1512.05751}}.

\bibitem{Bian:2015kjt}
L.~Bian, N.~Chen, D.~Liu, and J.~Shu, {\it {A hidden confining world on the 750
  GeV diphoton excess}},  \href{http://arxiv.org/abs/1512.05759}{{\tt
  arXiv:1512.05759}}.

\bibitem{Falkowski:2015swt}
A.~Falkowski, O.~Slone, and T.~Volansky, {\it {Phenomenology of a 750 GeV
  Singlet}},  \href{http://arxiv.org/abs/1512.05777}{{\tt arXiv:1512.05777}}.

\bibitem{Bai:2015nbs}
Y.~Bai, J.~Berger, and R.~Lu, {\it {A 750 GeV Dark Pion: Cousin of a Dark
  G-parity-odd WIMP}},  \href{http://arxiv.org/abs/1512.05779}{{\tt
  arXiv:1512.05779}}.

\bibitem{Dhuria:2015ufo}
M.~Dhuria and G.~Goswami, {\it {Perturbativity, vacuum stability and inflation
  in the light of 750 GeV diphoton excess}},
  \href{http://arxiv.org/abs/1512.06782}{{\tt arXiv:1512.06782}}.

\bibitem{Chakraborty:2015jvs}
I.~Chakraborty and A.~Kundu, {\it {Diphoton excess at 750 GeV: Singlet scalars
  confront naturalness}},  \href{http://arxiv.org/abs/1512.06508}{{\tt
  arXiv:1512.06508}}.

\bibitem{Wang:2015kuj}
F.~Wang, L.~Wu, J.~M. Yang, and M.~Zhang, {\it {750 GeV Diphoton Resonance, 125
  GeV Higgs and Muon g-2 Anomaly in Deflected Anomaly Mediation SUSY Breaking
  Scenario}},  \href{http://arxiv.org/abs/1512.06715}{{\tt arXiv:1512.06715}}.

\bibitem{Murphy:2015kag}
C.~W. Murphy, {\it {Vector Leptoquarks and the 750 GeV Diphoton Resonance at
  the LHC}},  \href{http://arxiv.org/abs/1512.06976}{{\tt arXiv:1512.06976}}.

\bibitem{Hernandez:2015ywg}
A.~E.~C. Hernandez and I.~Nisandzic, {\it {LHC diphoton 750 GeV resonance as an
  indication of $SU(3)_c\times SU(3)_L\times U(1)_X$ gauge symmetry}},
  \href{http://arxiv.org/abs/1512.07165}{{\tt arXiv:1512.07165}}.

\bibitem{Huang:2015rkj}
W.-C. Huang, Y.-L.~S. Tsai, and T.-C. Yuan, {\it {Gauged Two Higgs Doublet
  Model confronts the LHC 750 GeV di-photon anomaly}},
  \href{http://arxiv.org/abs/1512.07268}{{\tt arXiv:1512.07268}}.

\bibitem{Badziak:2015zez}
M.~Badziak, {\it {Interpreting the 750 GeV diphoton excess in minimal
  extensions of Two-Higgs-Doublet models}},
  \href{http://arxiv.org/abs/1512.07497}{{\tt arXiv:1512.07497}}.

\bibitem{Cvetic:2015vit}
M.~Cvetič, J.~Halverson, and P.~Langacker, {\it {String Consistency, Heavy
  Exotics, and the $750$ GeV Diphoton Excess at the LHC}},
  \href{http://arxiv.org/abs/1512.07622}{{\tt arXiv:1512.07622}}.

\bibitem{Cheung:2015cug}
K.~Cheung, P.~Ko, J.~S. Lee, J.~Park, and P.-Y. Tseng, {\it {A Higgcision study
  on the 750 GeV Di-photon Resonance and 125 GeV SM Higgs boson with the
  Higgs-Singlet Mixing}},  \href{http://arxiv.org/abs/1512.07853}{{\tt
  arXiv:1512.07853}}.

\bibitem{Zhang:2015uuo}
J.~Zhang and S.~Zhou, {\it {Electroweak Vacuum Stability and Diphoton Excess at
  750 GeV}},  \href{http://arxiv.org/abs/1512.07889}{{\tt arXiv:1512.07889}}.

\bibitem{Hall:2015xds}
L.~J. Hall, K.~Harigaya, and Y.~Nomura, {\it {750 GeV Diphotons: Implications
  for Supersymmetric Unification}},
  \href{http://arxiv.org/abs/1512.07904}{{\tt arXiv:1512.07904}}.

\bibitem{Wang:2015omi}
F.~Wang, W.~Wang, L.~Wu, J.~M. Yang, and M.~Zhang, {\it {Interpreting 750 GeV
  Diphoton Resonance in the NMSSM with Vector-like Particles}},
  \href{http://arxiv.org/abs/1512.08434}{{\tt arXiv:1512.08434}}.

\bibitem{Salvio:2015jgu}
A.~Salvio and A.~Mazumdar, {\it {Higgs Stability and the 750 GeV Diphoton
  Excess}},  \href{http://arxiv.org/abs/1512.08184}{{\tt arXiv:1512.08184}}.

\bibitem{Son:2015vfl}
M.~Son and A.~Urbano, {\it {A new scalar resonance at 750 GeV: Towards a proof
  of concept in favor of strongly interacting theories}},
  \href{http://arxiv.org/abs/1512.08307}{{\tt arXiv:1512.08307}}.

\bibitem{Cai:2015hzc}
C.~Cai, Z.-H. Yu, and H.-H. Zhang, {\it {The 750 GeV diphoton resonance as a
  singlet scalar in an extra dimensional model}},
  \href{http://arxiv.org/abs/1512.08440}{{\tt arXiv:1512.08440}}.

\bibitem{Bizot:2015qqo}
N.~Bizot, S.~Davidson, M.~Frigerio, and J.~L. Kneur, {\it {Two Higgs doublets
  to explain the excesses $pp\rightarrow \gamma\gamma(750\ {\rm GeV})$ and $h
  \to \tau^\pm \mu^\mp$}},  \href{http://arxiv.org/abs/1512.08508}{{\tt
  arXiv:1512.08508}}.

\bibitem{Hamada:2015skp}
Y.~Hamada, T.~Noumi, S.~Sun, and G.~Shiu, {\it {An O(750) GeV Resonance and
  Inflation}},  \href{http://arxiv.org/abs/1512.08984}{{\tt arXiv:1512.08984}}.

\bibitem{Kang:2015roj}
S.~K. Kang and J.~Song, {\it {Top-phobic heavy Higgs boson as the 750 GeV
  diphoton resonance}},  \href{http://arxiv.org/abs/1512.08963}{{\tt
  arXiv:1512.08963}}.

\bibitem{Jiang:2015oms}
Y.~Jiang, Y.-Y. Li, and T.~Liu, {\it {750 GeV Resonance in the Gauged
  $U(1)'$-Extended MSSM}},  \href{http://arxiv.org/abs/1512.09127}{{\tt
  arXiv:1512.09127}}.

\bibitem{Jung:2015etr}
S.~Jung, J.~Song, and Y.~W. Yoon, {\it {How Resonance-Continuum Interference
  Changes 750 GeV Diphoton Excess: Signal Enhancement and Peak Shift}},
  \href{http://arxiv.org/abs/1601.00006}{{\tt arXiv:1601.00006}}.

\bibitem{Gu:2015lxj}
J.~Gu and Z.~Liu, {\it {Running after Diphoton}},
  \href{http://arxiv.org/abs/1512.07624}{{\tt arXiv:1512.07624}}.

\bibitem{Goertz:2015nkp}
F.~Goertz, J.~F. Kamenik, A.~Katz, and M.~Nardecchia, {\it {Indirect
  Constraints on the Scalar Di-Photon Resonance at the LHC}},
  \href{http://arxiv.org/abs/1512.08500}{{\tt arXiv:1512.08500}}.

\bibitem{Ko:2016lai}
P.~Ko, Y.~Omura, and C.~Yu, {\it {Diphoton Excess at 750 GeV in leptophobic
  U(1)$^\prime$ model inspired by $E_6$ GUT}},
  \href{http://arxiv.org/abs/1601.00586}{{\tt arXiv:1601.00586}}.

\bibitem{Palti:2016kew}
E.~Palti, {\it {Vector-Like Exotics in F-Theory and 750 GeV Diphotons}},
  \href{http://arxiv.org/abs/1601.00285}{{\tt arXiv:1601.00285}}.

\bibitem{Karozas:2016hcp}
A.~Karozas, S.~F. King, G.~K. Leontaris, and A.~K. Meadowcroft, {\it {750 GeV
  Diphoton excess from $E_6$ in F-theory GUTs}},
  \href{http://arxiv.org/abs/1601.00640}{{\tt arXiv:1601.00640}}.

\bibitem{Bhattacharya:2016lyg}
S.~Bhattacharya, S.~Patra, N.~Sahoo, and N.~Sahu, {\it {750 GeV Di-photon
  excess at CERN LHC from a dark sector assisted scalar decay}},
  \href{http://arxiv.org/abs/1601.01569}{{\tt arXiv:1601.01569}}.

\bibitem{Cao:2016udb}
J.~Cao, L.~Shang, W.~Su, Y.~Zhang, and J.~Zhu, {\it {Interpreting the 750 GeV
  diphoton excess in the Minimal Dilaton Model}},
  \href{http://arxiv.org/abs/1601.02570}{{\tt arXiv:1601.02570}}.

\bibitem{Faraggi:2016xnm}
A.~E. Faraggi and J.~Rizos, {\it {The 750 GeV diphoton LHC excess and Extra Z's
  in Heterotic-String Derived Models}},
  \href{http://arxiv.org/abs/1601.03604}{{\tt arXiv:1601.03604}}.

\bibitem{Han:2016bvl}
X.-F. Han, L.~Wang, and J.~M. Yang, {\it {An extension of two-Higgs-doublet
  model and the excesses of 750 GeV diphoton, muon g-2 and $h\to\mu\tau$}},
  \href{http://arxiv.org/abs/1601.04954}{{\tt arXiv:1601.04954}}.

\bibitem{Kawamura:2016idj}
J.~Kawamura and Y.~Omura, {\it {Diphoton excess at 750 GeV and LHC constraints
  in models with vector-like particles}},
  \href{http://arxiv.org/abs/1601.07396}{{\tt arXiv:1601.07396}}.

\bibitem{King:2016wep}
S.~F. King and R.~Nevzorov, {\it {750 GeV Diphoton Resonance from Singlets in
  an Exceptional Supersymmetric Standard Model}},
  \href{http://arxiv.org/abs/1601.07242}{{\tt arXiv:1601.07242}}.

\bibitem{Nomura:2016rjf}
T.~Nomura and H.~Okada, {\it {Generalized Zee-Babu model with 750 GeV Diphoton
  Resonance}},  \href{http://arxiv.org/abs/1601.07339}{{\tt arXiv:1601.07339}}.

\bibitem{Harigaya:2016pnu}
K.~Harigaya and Y.~Nomura, {\it {A Composite Model for the 750 GeV Diphoton
  Excess}},  \href{http://arxiv.org/abs/1602.01092}{{\tt arXiv:1602.01092}}.

\bibitem{Han:2016fli}
C.~Han, T.~T. Yanagida, and N.~Yokozaki, {\it {Implications of the 750 GeV
  Diphoton Excess in Gaugino Mediation}},
  \href{http://arxiv.org/abs/1602.04204}{{\tt arXiv:1602.04204}}.

\bibitem{Hamada:2016vwk}
Y.~Hamada, H.~Kawai, K.~Kawana, and K.~Tsumura, {\it {Models of LHC Diphoton
  Excesses Valid up to the Planck scale}},
  \href{http://arxiv.org/abs/1602.04170}{{\tt arXiv:1602.04170}}.

\bibitem{Bae:2016xni}
K.~J. Bae, M.~Endo, K.~Hamaguchi, and T.~Moroi, {\it {Diphoton Excess and
  Running Couplings}},  \href{http://arxiv.org/abs/1602.03653}{{\tt
  arXiv:1602.03653}}.

\bibitem{Halverson:2013ska}
J.~Halverson, {\it {Anomaly Nucleation Constrains SU(2) Gauge Theories}},  {\em
  Phys. Rev. Lett.} {\bf 111} (2013), no.~26 261601,
  [\href{http://arxiv.org/abs/1310.1091}{{\tt arXiv:1310.1091}}].

\bibitem{Cvetic:2011iq}
M.~Cveti{\v c}, J.~Halverson, and P.~Langacker, {\it {Implications of String
  Constraints for Exotic Matter and Z' s Beyond the Standard Model}},  {\em
  JHEP} {\bf 11} (2011) 058.

\bibitem{Cvetic:2012kj}
M.~Cveti{\v c}, J.~Halverson, and H.~Piragua, {\it {Stringy Hidden Valleys}},
  {\em JHEP} {\bf 1302} (2013) 005.

\bibitem{Halverson:2014nwa}
J.~Halverson, N.~Orlofsky, and A.~Pierce, {\it {Vectorlike Leptons as the Tip
  of the Dark Matter Iceberg}},  {\em Phys. Rev.} {\bf D90} (2014), no.~1
  015002, [\href{http://arxiv.org/abs/1403.1592}{{\tt arXiv:1403.1592}}].

\bibitem{Martin:1997ns}
S.~P. Martin, {\it {A Supersymmetry primer}},
  \href{http://arxiv.org/abs/hep-ph/9709356}{{\tt hep-ph/9709356}}. [Adv. Ser.
  Direct. High Energy Phys.18,1(1998)].

\bibitem{Cvetic:2003ch}
M.~Cvetic and I.~Papadimitriou, {\it {Conformal field theory couplings for
  intersecting D-branes on orientifolds}},  {\em Phys. Rev.} {\bf D68} (2003)
  046001, [\href{http://arxiv.org/abs/hep-th/0303083}{{\tt hep-th/0303083}}].
  [Erratum: Phys. Rev.D70,029903(2004)].

\bibitem{Cvetic:1985fp}
M.~Cvetic and C.~R. Preitschopf, {\it {Heavy Families and $N=1$ Supergravity
  Within the Standard Model}},  {\em Nucl. Phys.} {\bf B272} (1986) 490.

\bibitem{Khachatryan:2015gza}
{\bf CMS} Collaboration, V.~Khachatryan et~al., {\it {Search for pair-produced
  vector-like B quarks in proton-proton collisions at $\sqrt{s}$ = 8 TeV}},
  \href{http://arxiv.org/abs/1507.07129}{{\tt arXiv:1507.07129}}.

\bibitem{Aad:2015kqa}
{\bf ATLAS} Collaboration, G.~Aad et~al., {\it {Search for production of
  vector-like quark pairs and of four top quarks in the lepton-plus-jets final
  state in $pp$ collisions at $\sqrt{s}=8$ TeV with the ATLAS detector}},  {\em
  JHEP} {\bf 08} (2015) 105, [\href{http://arxiv.org/abs/1505.04306}{{\tt
  arXiv:1505.04306}}].

\bibitem{Khachatryan:2015oba}
{\bf CMS} Collaboration, V.~Khachatryan et~al., {\it {Search for vector-like
  charge 2/3 T quarks in proton-proton collisions at $\sqrt(s)$ = 8 TeV}},
  {\em Phys. Rev.} {\bf D93} (2016), no.~1 012003,
  [\href{http://arxiv.org/abs/1509.04177}{{\tt arXiv:1509.04177}}].

\bibitem{Agashe:2014kda}
{\bf Particle Data Group} Collaboration, K.~A. Olive et~al., {\it {Review of
  Particle Physics (RPP)}},  {\em Chin.Phys.} {\bf C38} (2014) 090001. {\tt
  http://pdg.lbl.gov}.

\bibitem{Aad:2015dha}
{\bf ATLAS} Collaboration, G.~Aad et~al., {\it {Search for heavy lepton
  resonances decaying to a $Z$ boson and a lepton in $pp$ collisions at
  $\sqrt{s}=8$ TeV with the ATLAS detector}},  {\em JHEP} {\bf 09} (2015) 108,
  [\href{http://arxiv.org/abs/1506.01291}{{\tt arXiv:1506.01291}}].

\end{thebibliography}\endgroup

\end{document}